\documentclass[twocolumn]{svjour3}
\usepackage{amssymb}
\usepackage{amsmath}
\usepackage{bm}
\usepackage{dcolumn}
\usepackage{graphicx}

\usepackage{color}

\RequirePackage{fix-cm}
\smartqed
\begin{document}

\title{
Slow in-plane magnetoresistance oscillations in multiband quasi-two-dimensional metals
%Slow magnetoresistance oscillations as a tool to measure the electronic parameters in multiband metals and iron-based superconductors \\
%Slow oscillations of in-plane magnetoresistance in iron-based superconductors and other multi-band quasi-two-dimensional metals
}
\author{P.D. Grigoriev \and M.M. Korshunov \and T.I. Mogilyuk}
\institute{P.D. Grigoriev \at
           L.D. Landau Institute for Theoretical Physics, 142432, Chernogolovka, Russia;\
           \email{grigorev@itp.ac.ru} \\
           \emph{Temporal address:} Institut Laue-Langevin, BP 156, 6 rue Jules Horowitz, 38042 Grenoble Cedex 9, France \\
           \emph{Also at:} National University of Science and Technology ``MISiS'', Moscow 119049, Russia
           \and
           M.M. Korshunov \at L.V. Kirensky Institute of Physics, Krasnoyarsk 660036, Russia; \\
           Siberian Federal University, Svobodny Prospect 79, Krasnoyarsk 660041, Russia \\
           \email{mkor@iph.krasn.ru}
           \and
           T.I. Mogilyuk \at National Research Centre “Kurchatov Institute”, Moscow, Russia; \
           \email{5tarasm@gmail.com}
}
\date{Received: date / Accepted: date}
\maketitle

\begin{abstract}
Slow oscillations (SlO) of magnetoresistance is a convenient tool to measure electronic structure parameters in quasi-two-dimensional metals. We study the possibility to apply this method to multi-band conductors, e.g. to iron-based high-temperature superconducting materials. We show that SlO can be used to measure the interlayer transfer integral in multi-band conductors similar to single-band metals. In addition, the SlO allow to measure and compare the effective masses or the electron scattering rates in various bands.
\end{abstract}

% \PACS{74.70.Xa \and 74.20.Rp \and 74.62.En \and 74.25.F-}

%\authorrunning{P.D. Grigoriev, M.M. Korshunov} % if too long for running head


\section{Introduction}

Discovery of the superconductivity in iron-based materials~\cite{kamihara}
raised a question about the nature of the underlying electron pairing. Most
promising is the electronic mechanism of Cooper pair formation originating
from the dominating exchange of either spin or orbital fluctuations~\cite{reviews,Hirschfeld}. Former results in the extended $s$-wave order
parameter that change sign between electron and hole Fermi surface pockets,
the so-called $s_\pm$ state~\cite{Hirschfeld,Kuroki,Mazin,Chubukov,KorshunovUFN}. Orbital fluctuations
enhanced by the electron-phonon coupling may lead to the sign-preserving $s$-wave gap, the $s_{++}$ state~\cite{kontani,bang}.

Since electronic mechanisms of pairing involves particles near the Fermi
level, knowledge of the topology and details of the Fermi surface (FS) is
crucial. There are several experimental methods of determining it. Widely
used are angle-resolved photoemission spectroscopy (ARPES) and magnetic
quantum oscillations (MQO) measurements. ARPES provides a lot of valuable
information on the electronic structure~\cite{Kordyuk2011,Richard2011}
especially considering the quasi-two-dimensional nature of Fe-based
materials, but its surface sensitivity sometime may be a severe limitation.
In this respect, MQO are more reliable method of determining the bulk
properties. MQO measurements were performed on a number of Fe-based
materials, both pnictides and chalcogenides~\cite{Carrington2011}. In
particular, data are available for LaFePO~\cite{Coldea2008,Sugawara2008},
undoped 122 systems~\cite{Sebastian2008,Analytis2009,Analytis2009_2,Terashima2009,Sutherland2011},
BaFe$_{2}$(As$_{1-x}$P$_{x}$)$_{2}$~\cite{Shishido2010,Analytis2010}, KFe$_{2}$As$_{2}$~\cite{Terashima2010}, 111 systems LiFeP and LiFeAs~\cite{Putzke2012}, and 11 system FeSe~\cite{Watson2015}.

The iron-based superconducting materials, as well as most other high-$T_c$
superconductors, have a strong quasi-two-dimensional (Q2D) anisotropy of
electronic dispersion and conductivity. In the tight-binding
approximation the electronic dispersion of Q2D metals is given by
\begin{equation}
\epsilon _{3D}\left( \mathbf{k}\right) \approx \sum_{\eta }\epsilon _{\eta
}\left( \mathbf{k}_{||}\right) -2t_{z,\eta }\left( \mathbf{k}_{||}\right)
\cos (k_{z}d),  \label{E3D}
\end{equation}
where $\eta $ numerates different Fermi-surface pockets (or bands) with
in-plane dispersion $\epsilon _{\eta }\left( \mathbf{k}_{||}\right) $, $%
\mathbf{k}_{||}=\left\{ k_{x},k_{y}\right\} $ is the in-plane electron
momentum, $d$ is the interlayer lattice constant, and the interlayer
transfer integral $t_{z}$ is much less than the in-plane Fermi energy $%
E_{F\eta }=\mu _{\eta }$ of any band $\eta $. Below we assume that $t_{z}$\
is momentum-independent and the same for all bands $\eta $: $t_{z,\eta
}\left( \mathbf{k}_{||}\right) =t_{z}$.
%, because the value of $t_{z}$ is determined by the interlayer crystal-lattice potential of large energy scale.
Then the Fermi surface of each band is a cylinder with weak warping $%
\sim 4t_{z}/E_{F}\ll 1$. The MQO with such FS have two close fundamental
frequencies $F_{0}\pm \Delta F$. In a magnetic field $\boldsymbol{B}=B_{z}$
perpendicular to the conducting layers $F_{0}/B=\mu _{\eta }/\hbar \omega
_{c,\eta }$ and $\Delta F/B=2t_{z}/\hbar \omega _{c,\eta }$, where $\hbar
\omega _{c,\eta }=\hbar eB_{z}/m_{\eta }^{\ast }c$ is the distance between
the Landau levels (LL), sometimes called the cyclotron energy, and $m_{\eta
}^{\ast }$\ is an effective electron mass for this band $\eta $.

Magnetoresistance (MR) in layered Q2D conductors has interesting features,
which do not appear in 3D metals. At $2t_{z}<\mu $ the angular oscillations
of interlayer MR (AMRO) have been observed in many layered organic metals
(see, e.g., Refs.~\cite{MQORev,OMRev,MarkReview2004,KartPeschReview} for
reviews) and in some cuprate high-$T_c$ superconductors~\cite{HusseyNature2003}, which was interpreted as a signature of a well-defined quasi-2D Fermi
surface in these materials. For an isotropic in-plane electron dispersion
AMRO are qualitatively described by the renormalization of the interlayer
transfer integral:\cite{Kur}
\begin{equation}
t_{z}=t_{z}\left( \theta \right) =t_{z}\left( 0\right) J_{0}\left(
k_{F}d\tan \theta \right) ,  \label{tz}
\end{equation}
where $J_{0}\left( x\right) $ is the Bessel's function, $p_{F}=\hbar k_{F}$
is the in-plane Fermi momentum, and $\theta $ is the angle between the
magnetic field $\boldsymbol{B}$ and the normal to conducting layers.

At $t_{z}\sim \hbar \omega _{c}$ several additional qualitative features of
MR appear. For example, the strong monotonic growth of interlayer MR $R_{zz}(B_{z})$ was observed in various Q2D metals~\cite{Coldea,PesotskiiJETP95,Zuo1999,W3,W4,Incoh2009,Kang,Wosnitza2002,SO,WIPRB2012}
and recently theoretically explained~\cite{WIPRB2012,WIPRB2011,WIPRB2013,GG2014}. At $t_{z}\gtrsim \hbar \omega _{c}$
the MR acquires the so-called slow oscillations~\cite{SO,Shub} and the phase
shift of beats~\cite{PhSh,Shub}. These two effects appear in the higher
orders in $\hbar \omega_{c}/t_{z}$ and, therefore, are missed in the
standard 3D theory of MR~\cite{Abrik,Shoenberg,Ziman}.

These slow oscillations (SlO) originate from the finite interlayer hopping $t_{z}$ contrary to usual MQO with low frequency, originating from small FS
pockets. The product of oscillations with two close frequencies $F_{0}\pm
\Delta F$ gives oscillations with frequency $2\Delta F$:
%$\cos \left( F_{0}+\Delta F\right) \cos \left( F_{0}-\Delta F\right) = \left[ \cos \left( 2F_{0}\right) +\cos \left( 2\Delta F\right) \right] /2$.
\begin{equation}
\cos \left(F_{0}+\Delta F\right) \cos \left( F_{0}-\Delta F\right) = \frac{ \cos \left(
2F_{0}\right) + \cos \left( 2\Delta F\right)}{2} .
\end{equation}
The conductivity,
being a non-linear function of the oscillating electronic density of states
(DoS) and of the diffusion coefficient, has SlO with frequency $2 \Delta
F\propto t_{z}$, while the magnetization, being a linear functional of DoS,
does not show SlO~\cite{SO,Shub}. The SlO have many interesting and useful
features as compared to the quantum oscillations. First, they survive at
much higher temperature than MQO, because, contrary to MQO, they are not
suppressed by the temperature smearing of Fermi distribution function.
Second, they are not sensitive to a long-range disorder, which damps the
fast MQO similarly to finite temperature due to a spatial variation of the
Fermi energy. Third, the SlO allow to measure the interlayer transfer
integral $t_{z}$ and the in-plane Fermi momentum $p_{F}\equiv \hbar k_{F}$.
These features make the SlO to be a useful tool to study the electronic
properties of Q2D metals. Almost 30 years since their discovery~\cite{Mark88}
and more than 10 years after their explanation~\cite{SO,Shub} the SlO where
investigated only for the \textit{interlayer} conductivity $\sigma
_{zz}\left( B\right) $, when the current and the magnetic field are both
applied perpendicularly to the 2D layers, and only in organic compounds. The
SlO were shown to be a useful and very accurate tool to measure the
interlayer transfer integral $t_{z}$. In addition, the SlO allow to obtain
information about the in-plane Fermi momentum $k_{F}$\ and even about the
type of disorder, as short- or long-range disorder~\cite{SO,Shub}. Later it
was realized that the monocrystals of most layered Q2D compounds, including pnictide
high-temperature superconductors, as a rule, have the shape of very thin
flakes for which the correct measurements of the interlayer conductivity is
very difficult, especially in the case of good metallic properties of
studied compounds. Recently, the first measurements and qualitative analysis
of SlO of the intralayer (in-plane) conductivity in the non-organic layered
Q2D rare-earth tritelluride compounds $R$Te$_{3}$ ($R=$Gd and Tb) was
reported~\cite{SlowPRB2015}. From these experimental data for the first time
in these strongly anisotropic Q2D conductors the authors obtained the value
of the interlayer transfer integral $t_{z}$ and estimated the in-plane Fermi
momentum after the FS reconstruction due to the double charge-density-wave
superstructure~\cite{SlowPRB2015}. Thus, the slow oscillations of MR proved
to be a powerful technique to explore the electronic structure of various
compounds. In this report we investigate the possibility of its application
to the multiband systems like iron-based high-$T_c$ superconducting materials.

Contrary to the situation in the strongly correlated high-$T_{c}$ cuprates,
consensus between electronic band structure calculations in the density
functional theory (DFT), ARPES, QO, and Compton scattering~\cite{Utfeld2010}
has been promptly established, so that the in-plane electron dispersion $\epsilon _{\eta }\left( \mathbf{k}_{||}\right)$ for each band $\eta$ is
known. The gross feature is that excluding the cases of extreme hole and
electron dopings, the in-plane FS in Fe-based materials consists of two or
three hole pockets around the $\Gamma =(0,0)$ point and two electron pockets
around the $M=(\pi ,\pi )$ point in the 2-Fe Brillouin zone corresponding to
the crystallographic unit cell. The scattering between these two groups of
FSs believed to be responsible for the stripe antiferromagnetic order in
undoped materials and for the spin fluctuation mediated pairing in doped
compounds. Thus the two-band model is the minimal model capturing the basic
yet essential physics of pnictides and chalcogenides (see discussion in Ref.~\cite{Hirschfeld}). Below we generalize the qualitative study of SlO of
intralayer MR in Ref.~\cite{SlowPRB2015} from the one-band to a multi-band
model, assuming that there are $\lambda $ different %types of
bands.

\section{Calculations}

According to Eq. (90.5) of Ref.~\cite{LL10} the intralayer conductivity at
finite temperature is given by~\cite{LL10}
\begin{equation}
\sigma _{yy}=\int d\varepsilon \,\left[ -n_{F}^{\prime }(\varepsilon )\right]
\,\sigma _{yy}(\varepsilon ),  \label{sigmayy}
\end{equation}
where the derivative of the Fermi distribution function $n_{F}^{\prime}(\varepsilon )=-1/\{4T\cosh ^{2}\left[ (\varepsilon -\mu )/2T\right] \}$,
and the zero-temperature electron conductivity at energy $\varepsilon $ is
\begin{equation}
\sigma _{yy}(\varepsilon )=\sum_{\eta }\sigma _{yy,\eta }(\varepsilon
)=\sum_{\eta }e^{2}g_{\eta }\left( \varepsilon \right) D_{y,\eta }\left(
\varepsilon \right) .  \label{s1}
\end{equation}
Here $g_{\eta }\left( \varepsilon \right) $ is the DoS and $D_{y,\eta}\left( \varepsilon \right) $ is the diffusion coefficient along y-axis of
electrons from the band $\eta $. It is convenient to use the harmonic
expansion for the oscillating DoS $g_{\eta }\left( \varepsilon \right) $.
Below we will need only the first terms in this harmonic series, which at
finite $t_{z}\sim \hbar \omega _{c}$ are given by~\cite{Champel2001,ChampelMineev,Shub}
\begin{equation}
g_{\eta }\left( \varepsilon \right) \approx g_{0,\eta }\left[ 1-2\cos \left(
\frac{2\pi \varepsilon }{\hbar \omega _{c,\eta }}\right) J_{0}\left( \frac{4\pi t_{z}}{\hbar \omega _{c,\eta }}\right) R_{D,\eta }\right] ,  \label{ge1}
\end{equation}%
where $g_{0,\eta }=m_{\eta }^{\ast }/\pi \hbar ^{2}d$ is the DoS per two
spin components at the Fermi level from the band $\eta $ in the absence of magnetic field, $J_{0}\left( x\right) $ is the Bessel's function, $R_{D,\eta }\approx \exp \left[ -\pi /\omega _{c,\eta }\tau _{0,\eta }\right] $ is the
Dingle factor~\cite{Dingle,Bychkov}, $\tau _{0,\eta }$ is the
electron mean free time without magnetic field, which for scattering by point-like impurities depends
only on the total DoS and not on the band index $\eta $: $\tau _{0,\eta }=\tau _{0}=\hbar /2\Gamma _{0}$, where $\Gamma _{0}$
is the LL broadening.

The calculation of the diffusion coefficient $D_{y}\left( \varepsilon
\right) $ is less trivial and requires to specify the model. At $\mu \gg
\hbar \omega _{c}$ the quasi-classical approximation is applicable. In an
ideal crystal in a magnetic field $\boldsymbol{B}$ the electrons move along
the cyclotron orbits with a fixed center and the Larmor radius of band $\eta$, $R_{L,\eta} = p_{F,\eta }c/eB_{z}$. Without scattering the electron
diffusion in the direction perpendicular to $\boldsymbol{B}$ is absent. The
electron-electron (e-e) interaction in the absence of magnetic field and of
umklapp processes does not change the total electron momentum and, hence,
does not change electric conductivity, though in combination with disorder,
the e-e interaction leads to substantial corrections to conductivity~\cite{AABook}. Scattering by impurities changes the electronic states and leads
to the electron diffusion perpendicular to magnetic field, and we take into
account only this mechanism of the in-plane electron diffusion in
perpendicular magnetic field. For simplicity, we consider only the
scattering by short-range impurities, described by the $\delta $-function
potential: $V_{i}\left( r\right) = U\delta ^{3}\left( r-r_{i}\right) $.
Scattering by impurities is elastic, i.e. it conserves the electron energy $\varepsilon$, but the quantum numbers of electron states may change. The
matrix element of impurity scattering is given by
\begin{equation}
T_{mm^{\prime }}=\Psi _{m^{\prime }}^{\ast }\left( r_{i}\right) U\Psi
_{m}\left( r_{i}\right) ,  \label{Tm}
\end{equation}
where $\Psi _{m}\left( r\right) $ is the electron wave function in the state
$m$. During each scattering, the typical change $\Delta y=\Delta
P_{x}c/eB_{z}$ of the mean electron coordinate $y_{0}$ perpendicular to $\boldsymbol{B}$ is of the order of $R_{L,\eta }$, because for larger $\Delta
y\gg R_{L}$ the matrix element in Eq.~(\ref{Tm}) is exponentially small
because of small overlap of the electron wave functions $\Psi _{m^{\prime
}}^{\ast }\left( r_{i}\right) \Psi _{m}\left( r_{i}\right) \sim \Psi
_{m}^{\ast }\left( r_{i}+\Delta y\right) \Psi _{m}\left( r_{i}\right)$~\cite{CommentDecay}. The diffusion coefficient for the band $\eta $ is
approximately given by
\begin{equation}
D_{y,\eta }\left( \varepsilon \right) \approx \left\langle \left( \Delta
y\right) ^{2}\right\rangle _{\eta }~/2\tau _{\eta }\left( \varepsilon
\right) ,  \label{D1}
\end{equation}%
where $\tau _{\eta }\left( \varepsilon \right) $ is the energy-dependent
electron mean scattering time by impurities, and the angular brackets in Eq.~(\ref{D1}) mean averaging over impurity scattering events. In the Born
approximation, the mean scattering rate on point-like impurities is
independent on the band $\eta $ and given by
\begin{equation}
1/\tau _{\eta }\left( \varepsilon \right) =1/\tau \left( \varepsilon \right)
=2\pi n_{i}U^{2}g\left( \varepsilon \right) ,  \label{tau}
\end{equation}%
where $n_{i}$ is the impurity concentration and $g\left( \varepsilon \right)
\equiv \sum_{\eta }g_{\eta }\left( \varepsilon \right) $ is the total DoS.
This scattering rate has MQO, which are reduced as compared to those of the
DoS in Eq.~(\ref{ge1}), because MQO of the DoS from different bands have
different frequencies and partially cancel each other. Indeed, if one takes
the total number of bands $\lambda >1$ and the same average DoS from each
band: $g_{0,\eta }=g_{0}$, one obtains
\begin{equation}
\frac{g\left( \varepsilon \right) }{\lambda g_{0}}\approx 1-\sum_{\eta }%
\frac{2}{\lambda }\cos \left( \frac{2\pi \varepsilon }{\hbar \omega _{c,\eta
}}\right) J_{0}\left( \frac{4\pi t_{z}}{\hbar \omega _{c,\eta }}\right)
R_{D,\eta }.  \label{gtot}
\end{equation}

The MQO of $\left\langle \left( \Delta y\right) ^{2}\right\rangle _{\eta
}\approx R_{L,\eta }^{2}$ are, usually, weaker than MQO of $g_{\eta }\left(
\varepsilon \right) $, and in 3D metals they are neglected~\cite{LL10}. Then
\begin{equation}
D_{y,\eta }\left( \varepsilon \right) \approx R_{L,\eta ~}^{2}/2\tau \left(
\varepsilon \right) \propto g\left( \varepsilon \right) .  \label{D2}
\end{equation}%
\ However, in Q2D metals, when $t_{z}\sim \hbar \omega _{c}$, the MQO of $\left\langle \left( \Delta y\right) ^{2}\right\rangle $ can be of the same
order as MQO of the DoS. Moreover, they are not suppressed as strongly by
the averaging over various bands $\eta $, as $g\left( \varepsilon \right) $
is. Therefore, instead of Eq.~(\ref{D2}) at $R_{D}\ll 1$ one has
\begin{eqnarray}
\frac{D_{y,\eta }\left( \varepsilon \right) }{D_{0,\eta }} &\approx& 1-2\alpha
_{\eta }\cos \left( \frac{2\pi \varepsilon }{\hbar \omega _{c,\eta }}\right)
J_{0}\left( \frac{4\pi t_{z}}{\hbar \omega _{c,\eta }}\right) R_{D,\eta }
\label{D3} \\
&-&\sum_{\eta ^{\prime }\neq \eta }2\beta _{\eta \eta ^{\prime }}\cos \left(
\frac{2\pi \varepsilon }{\hbar \omega _{c,\eta ^{\prime }}}\right)
J_{0}\left( \frac{4\pi t_{z}}{\hbar \omega _{c,\eta ^{\prime }}}\right)
R_{D,\eta ^{\prime }},  \nonumber
\end{eqnarray}%
where $D_{0,\eta }\approx R_{L,\eta }^{2}/2\tau _{\eta }$, and the numbers $%
\alpha _{\eta }\sim 1$ and $\beta _{\eta \eta ^{\prime }}\sim 1/\lambda $.
Combining Eqs. (\ref{s1}), (\ref{ge1}) and (\ref{D3}) one obtains%
\begin{eqnarray}
\sigma _{yy}(\varepsilon ) &=& e^{2}\sum_{\eta }g_{0,\eta }D_{0,\eta }\times
\label{syy2} \\
&\times& \left[ 1-2\cos \left( \frac{2\pi \varepsilon }{\hbar \omega _{c,\eta }%
}\right) J_{0}\left( \frac{4\pi t_{z}}{\hbar \omega _{c,\eta }}\right)
R_{D,\eta }\right]   \nonumber \\
&\times& \left[ 1-2\alpha _{\eta }\cos \left( \frac{2\pi \varepsilon }{\hbar
\omega _{c,\eta }}\right) J_{0}\left( \frac{4\pi t_{z}}{\hbar \omega
_{c,\eta }}\right) R_{D,\eta }\right. \nonumber \\
&-&\left.\sum_{\eta ^{\prime }\neq \eta }2\beta _{\eta \eta ^{\prime }}\cos
\left( \frac{2\pi \varepsilon }{\hbar \omega _{c,\eta ^{\prime }}}\right)
J_{0}\left( \frac{4\pi t_{z}}{\hbar \omega _{c,\eta ^{\prime }}}\right)
R_{D,\eta ^{\prime }}\right] .  \nonumber
\end{eqnarray}%
The slow oscillations arise from the product of second terms in both square brackets %$\cos \left( 2\pi \varepsilon /\hbar \omega _{c,\eta }\right) \cos \left( 2\pi \varepsilon /\hbar \omega _{c,\eta}\right)$
with the same cyclotron frequency $\omega_{c,\eta }=\omega_{c,\eta ^{\prime }}$, i.e. %from
$\eta =\eta ^{\prime }$,
because only these terms give the energy-independent term $1/2$, which is
not affected by the averaging over $\varepsilon $: $\cos ^{2}\left( 2\pi
\varepsilon /\hbar \omega _{c,\eta }\right) =\left[ 1+\cos \left( 4\pi
\varepsilon /\hbar \omega _{c,\eta }\right) \right] /2$. Hence, the
classical (monotonic + slow oscillating) part of $\sigma _{yy}(B)$ is
obtained by collecting all leading energy independent terms in Eq.~(\ref{sigmayy}) with subsequent trivial integration over $\varepsilon $ after
substitution to Eq.~(\ref{sigmayy}),
\begin{equation}
\sigma _{yy}^{SlO}(B)\approx e^{2}\sum_{\eta }g_{0,\eta }D_{0,\eta }\left[
1+2\alpha _{\eta }J_{0}^{2}\left(\frac{4\pi t_{z}}{\hbar \omega _{c,\eta }}\right)
R_{D,\eta }^{2}\right] .  \label{SO}
\end{equation}%
The other cross products in Eq.~(\ref{syy2}) give MQO, i.e. the $\varepsilon
$-dependent terms $\propto \cos \left( 2\pi \varepsilon /\hbar \omega
_{c,\eta ^{\prime }}\right) $, which after temperature smearing in Eq.~(\ref{sigmayy}) acquire the usual temperature damping factor of MQO:
\begin{equation}
R_{T,\eta }=\left( 2\pi ^{2}k_{B}T/\hbar \omega _{c,\eta }\right) /\sinh
\left( 2\pi ^{2}k_{B}T/\hbar \omega _{c,\eta }\right) .  \label{RT}
\end{equation}%
On contrary, the SlO in Eq.~(\ref{SO}) are not damped by temperature within
our model.

Approximately, one can use the asymptotic expansion of the Bessel function
in Eq. (\ref{SO}) for large values of the argument: $J_{0}(x)\approx \sqrt{2/\pi x}\cos \left( x-\pi /4\right)$, $x\gg 1$. Then, after introducing
the frequencies of SlO,
\begin{equation}
F_{SlO,\eta }=4t_{z}B/\hbar \omega _{c,\eta },  \label{Fslow0}
\end{equation}
Eq.~(\ref{SO}) simplifies to
\begin{eqnarray}
\sigma _{yy}^{SlO}(B) &\approx& e^{2}\sum_{\eta }g_{0,\eta }D_{0,\eta }\times
\label{SOa} \\
&\times& \left[ 1+\frac{\alpha \hbar \omega _{c,\eta }}{2\pi ^{2}t_{z}}\sin
\left( \frac{2\pi F_{SlO,\eta }}{B}\right) R_{D,\eta }^{2}\right] .  \nonumber
\end{eqnarray}

In tilted magnetic field at constant $\left\vert \boldsymbol{B}\right\vert $, $\omega _{c} \propto \cos \theta $ and $t_{z}$ changes according to Eq.~(\ref{tz}). Then the frequencies of SlO will depend on tilt angle $\theta $
of magnetic field (with respect to the normal to conducting layers) as
\begin{equation}
F_{SlO,\eta }\left( \theta \right) /F_{SlO,\eta }\left( 0\right)
=J_{0}\left( k_{F,\eta }d\tan \theta \right) /\cos \left( \theta \right) .
\label{Fslow}
\end{equation}%
Note that this dependence is non-monotonic and crucially different from the
angular dependence of MQO frequencies, given by the simple cosine law: $F_{MQO}\left( \theta \right) / F_{MQO}\left( 0\right) = 1/\cos \left( \theta
\right) $.

\section{Discussion and conclusions}

As one can see by comparing Eq.~(\ref{SOa}) with the results
of Ref.~\cite{SlowPRB2015}, for multiband conductors both the slow
oscillations and MQO of magnetoresistance are damped in their relative amplitude by the
factor $\sim 1/\lambda$ as compared to single-band conductors, where $\lambda$ is the number of different bands. The origin is the different contributions to the sum over $\eta$ from the first (unity) and the second (band-dependent) terms in the square brackets. This is similar to the relative damping of the MQO of the DoS in Eq. (\ref{gtot}). Nevertheless, the SlO can be observed and used to extract the parameters of electronic structure from experimental data.

As one can see from Eq.~(\ref{SOa}), the slow oscillations of MR in
multi-band conductors are in most aspects similar to SlO in single-band
conductors, studied in Refs.~\cite{SO,Shub,SlowPRB2015}. Each frequency of
SlO corresponds to a particular band $\eta $ and can be used to extract
electronic parameters of this band. For the case of in-plane momentum-independent interlayer hopping $t_z$,
if the cyclotron mass or Landau-level
separation is known for at least one band, i.e. from the temperature
dependence of MQO amplitude, the frequency $F_{SlO,\eta }$ of slow
oscillations of magnetoresistance for this band gives the value of the
interlayer transfer integral $t_{z}$ according to Eq.~(\ref{Fslow0}). The
angular dependence of the SlO frequency has a non-monotonic angular
dependence given by Eq.~(\ref{Fslow}), which allows to extract the Fermi
momentum $k_{F,\eta }$ for this particular band as function of the azimuth
angle $\phi $ from experimental data on SlO. Since the interlayer transfer
integral is the same for all bands, the measured ratios of the SlO
frequencies for various bands $\eta $ give the ratios of their effective
(cyclotron) masses $m_{\eta }^{\ast }$, which allows to determine $m_{\eta
}^{\ast }$ for all bands if $m_{\eta }^{\ast }$ is known for at least one
band. This application of SlO was absent for single-band metals, being new
for the multi-band conductors. This \ application is very helpful, because
the temperature dependence of the MQO amplitudes cannot always be clearly
fitted by the Lifshitz-Kosevich formula and by Eq.~(\ref{RT}) for all
observed frequencies. If there is an independent way to determine $t_{z}$,
the SlO give an alternative way to determine all effective masses $m_{\eta
}^{\ast }$. The damping of slow oscillations, determined only by the Dingle factor,
can be used to compare the Dingle temperatures and, therefore, the scattering amplitudes
for different bands.

To summarize, in this paper we have shown the possibility of using rather
new phenomenon, namely, the slow oscillations of magnetoresistance, to
measure the parameters of electronic structure of multi-band quasi-two-dimensional conductors. The
application of this method to multi-band conductors has some specific
features, absent for single-band metals, which allow to extract and compare
the electronic parameters of different bands. We believe, that this
technique can be used to measure the interlayer transfer integral $t_{z}$
and other important parameters in iron-based pnictides and chalcogenides, MgB$_2$, Sr$_2$RuO$_4$, and in a variety of multiband conductors and superconductors.
%high-$T_c$ superconducting materials.

\begin{acknowledgements}
We are grateful to V.M. Pudalov, I.A. Nekrasov and S.G. Ovchinnikov for useful discussions. We acknowledge the partial support by RFBR (Grants 13-02-01395 and 13-02-00178) and the President Grant for Government Support of the Leading Scientific Schools of the Russian Federation (NSh-2886.2014.2).
\end{acknowledgements}


\begin{thebibliography}{99}
\bibitem{kamihara} Y. Kamihara, T. Watanabe, M. Hirano, and H. Hosono, J.
Am. Chem. Soc. \textbf{130}, 3296 (2008).

\bibitem{reviews} See reviews M.V. Sadovskii, Physics-Uspekhi \textbf{51},
1201 (2008); J. Paglione and R.L. Greene, Nat. Phys. \textbf{6}, 645 (2010);
D.C. Johnston, Adv. Phys. \textbf{59}, 803 (2010); H.H. Wen and S. Li, Annu.
Rev. Cond. Matter Phys. \textbf{2}, 121 (2011); G.R. Stewart, Rev. Mod.
Phys. \textbf{83}, 1589 (2011); E. Dagotto, Rev. Mod. Phys. \textbf{85}, 849
(2013), and refernces therein.

\bibitem{Hirschfeld} P.J. Hirschfeld, M.M. Korshunov, and I.I. Mazin, Rep.
Prog. Phys. \textbf{74}, 124508 (2011).

\bibitem{Kuroki} K. Kuroki, S. Onari, R. Arita, H. Usui, Y. Tanaka, H.
Kontani, and H. Aoki, Phys. Rev. Lett. \textbf{101}, 087004 (2008).

\bibitem{Mazin} I.I. Mazin, Nature (London) \textbf{464}, 183 (2010).

\bibitem{Chubukov} A.V. Chubukov, Annu. Rev. Cond. Matter Phys. \textbf{3},
57 (2012).

\bibitem{KorshunovUFN} M.M. Korshunov, Physics-Uspekhi \textbf{57}, 813
(2014).

\bibitem{kontani} H. Kontani and S. Onari, Phys. Rev. Lett. \textbf{104},
157001 (2010).

\bibitem{bang} Y. Bang, H.-Y. Choi, and H. Won, Phys. Rev. B \textbf{79},
054529 (2009).

% ARPES review

\bibitem{Kordyuk2011} A.A. Kordyuk, V.B. Zabolotnyy, D.V. Evtushinsky, A.N.
Yaresko, B. Buechner, and S.V. Borisenko, J. Supercond. Nov. Magn. \textbf{26%
}, 2837 (2013).

% ARPES review

\bibitem{Richard2011} P. Richard, T. Sato, K. Nakayama, T. Takahashi, and H.
Ding, Rep. Prog. Phys. \textbf{74}, 124512, (2011).

% review of quantum oscillations in FeBS

\bibitem{Carrington2011} A. Carrington, Rep. Prog. Phys. \textbf{74}, 124507
(2011).

% LaFePO (Tc ~ 6K) de Haas–van Alphen (dHvA) effect

\bibitem{Coldea2008} A.I. Coldea, J.D. Fletcher, A. Carrington, J.G.
Analytis, A.F. Bangura, J.-H. Chu, A.S. Erickson, I.R. Fisher, N.E. Hussey,
and R.D. McDonald, Phys. Rev. Lett. \textbf{101}, 216402 (2008).

% LaFePO de Haas–van Alphen Effect

\bibitem{Sugawara2008} H. Sugawara, R. Settai, Y. Doi, H. Muranaka, K.
Katayama, H. Yamagami, and Y. Onuki, J. Phys. Soc. Jpn. \textbf{77}, 113711
(2008).

% SrFe2As2 quantum oscillations

\bibitem{Sebastian2008} S.E. Sebastian, J. Gillett, N. Harrison, P.H.C. Lau,
D.J. Singh, C.H. Mielke, and G.G. Lonzarich, J. Phys.: Condens. Matter
\textbf{20}, 422203 (2008).

% BaFe2As2 de Haas–van Alphen Effect

\bibitem{Analytis2009} J.G. Analytis, R.D. McDonald, J.-H. Chu, S.C. Riggs,
A.F. Bangura, C. Kucharczyk, M. Johannes, and I.R. Fisher, Phys. Rev. B
\textbf{80}, 064507 (2009).

% SrFe2P2 de Haas–van Alphen effect

\bibitem{Analytis2009_2} J.G. Analytis, C.M.J. Andrew, A.I. Coldea, A.
McCollam, J.-H. Chu, R.D. McDonald, I.R. Fisher, and A. Carrington, Phys.
Rev. Lett. \textbf{103}, 076401 (2009).

% BaNi2P2 de Haas–van Alphen Effect

\bibitem{Terashima2009} T. Terashima, M. Kimata, H. Satsukawa, A. Harada, K.
Hazama, M. Imai, S. Uji, H. Kito, A. Iyo, H. Eisaki, and H. Harima, J. Phys.
Soc. Jpn. \textbf{78}, 033706 (2009).

% SrFe2As2 Shubnikov–de Haas

\bibitem{Sutherland2011} M. Sutherland, D.J. Hills, B.S. Tan, M.M.
Altarawneh, N. Harrison, J. Gillett, E.C.T. O'Farrell, T.M. Benseman, I.
Kokanovic, P. Syers, J.R. Cooper, and S.E. Sebastian, Phys. Rev. B \textbf{84%
}, 180506(R) (2011).

% BaFe2(As1-xPx)2 de Haas–van Alphen effect

\bibitem{Shishido2010} H. Shishido, A.F. Bangura, A.I. Coldea, S. Tonegawa,
K. Hashimoto, S. Kasahara, P.M.C. Rourke, H. Ikeda, T. Terashima, R. Settai,
Y. Onuki, D. Vignolles, C. Proust, B. Vignolle, A. McCollam, Y. Matsuda, T.
Shibauchi, and A. Carrington, Phys. Rev. Lett. \textbf{104}, 057008 (2010).

% BaFe2(As1−xPx)2 de Haas–van Alphen Effect

\bibitem{Analytis2010} J.G. Analytis, J.-H. Chu, R.D. McDonald, S.C. Riggs,
and I.R. Fisher, Phys. Rev. Lett. \textbf{105}, 207004 (2010).

% KFe2As2 de Haas–van Alphen Effect

\bibitem{Terashima2010} T. Terashima, M. Kimata, N. Kurita, H. Satsukawa, A.
Harada, K. Hazama, M. Imai, A. Sato, K. Kihou, C.-H. Lee, H. Kito, H.
Eisaki, A. Iyo, T. Saito, H. Fukazawa, Y. Kohori, H. Harima, and S. Uji, J.
Phys. Soc. Jpn. \textbf{79}, 053702 (2010).

% LiFeP and LiFeAs de Haas–van Alphen

\bibitem{Putzke2012} C. Putzke, A.I. Coldea, I. Guillam\'{o}n, D. Vignolles,
A. McCollam, D. LeBoeuf, M.D. Watson, I.I. Mazin, S. Kasahara, T. Terashima,
T. Shibauchi, Y. Matsuda, and A. Carrington, Phys. Rev. Lett. \textbf{108},
047002 (2012).

% FeSe Quantum oscillations

\bibitem{Watson2015} M.D. Watson, T.K. Kim, A.A. Haghighirad, N.R. Davies,
A. McCollam, A. Narayanan, S.F. Blake, Y.L. Chen, S. Ghannadzadeh, A.J.
Schofield, M. Hoesch, C. Meingast, T. Wolf, and A.I. Coldea, Phys. Rev. B
\textbf{91}, 155106 (2015).

% optimally doped Ba(Fe1−xCox)2As2 Compton scattering

\bibitem{Utfeld2010} C. Utfeld, J. Laverock, T.D. Haynes, S.B. Dugdale, J.A.
Duffy, M.W. Butchers, J.W. Taylor, S.R. Giblin, J.G. Analytis, J.-H. Chu,
I.R. Fisher, M. Itou, and Y. Sakurai, Phys. Rev. B \textbf{81}, 064509
(2010).

\bibitem{MQORev} J. Wosnitza, \textit{Fermi Surfaces of Low-Dimensional
Organic Metals and Superconductors} (Springer-Verlag, Berlin, 1996); J.
Singleton, Rep. Prog. Phys. \textbf{63}, 1111 (2000).

\bibitem{OMRev} T.~Ishiguro, K.~Yamaji and G.~Saito, \emph{Organic
Superconductors}, 2nd Edition, Springer-Verlag, Berlin, 1998; \textit{The
Physics of Organic Superconductors and Conductors}, ed. by A. G. Lebed
(Springer Series in Materials Science, V. 110; Springer Verlag Berlin
Heidelberg 2008).

\bibitem{MarkReview2004} M.V. Kartsovnik, Chem. Rev. \textbf{104}, 5737
(2004).

\bibitem{KartPeschReview} M. V. Kartsovn\u{\i}k and V. G. Peschansky, Low
Temp. Phys. \textbf{31}, 185 (2005) [Fiz. Nizk. Temp. \textbf{31}, 249
(2005)].

\bibitem{HusseyNature2003} N. E. Hussey, M. Abdel-Jawad, A. Carrington, A.
P. Mackenzie and L. Balicas, Nature \textbf{425}, 814 (2003).

\bibitem{Kur} Yasunari Kurihara, J. Phys. Soc. Jpn. \textbf{61}, 975 (1992).

\bibitem{Coldea} A. I. Coldea, A. F. Bangura, J. Singleton, A. Ardavan, A.
Akutsu-Sato, H. Akutsu, S. S. Turner, and P. Day, Phys. Rev. B \textbf{69},
085112 (2004).

\bibitem{PesotskiiJETP95} R.B. Lyubovskii, S.I. Pesotskii, A. Gilevskii and
R.N. Lyubovskaya, JETP \textbf{80}, 946 (1995) [Zh. Eksp. Teor. Fiz. \textbf{%
107}, 1698 (1995)].

\bibitem{Zuo1999} F. Zuo, X. Su, P. Zhang, J. S. Brooks, J. Wosnitza, J. A.
Schlueter, Jack M. Williams, P. G. Nixon, R. W. Winter, and G. L. Gard,
Phys. Rev. B \textbf{60}, 6296 (1999).

\bibitem{W3} J. Hagel, J. Wosnitza, C. Pfleiderer, J. A. Schlueter, J.
Mohtasham, and G. L. Gard, Phys. Rev. B \textbf{68}, 104504 (2003).

\bibitem{W4} J.Wosnitza, Journal of Low Temperature Physics 146, 641 (2007).

\bibitem{Incoh2009} M. V. Kartsovnik, P. D. Grigoriev, W. Biberacher, and N.
D. Kushch, Phys. Rev. B \textbf{79}, 165120 (2009).

\bibitem{Kang} W. Kang, Y. J. Jo, D. Y. Noh, K. I. Son, and Ok-Hee Chung,
Phys. Rev. B \textbf{80}, 155102 (2009).

\bibitem{Wosnitza2002} J.Wosnitza, J. Hagel, J. S. Qualls, J. S. Brooks, E.
Balthes, D. Schweitzer, J. A. Schlueter, U. Geiser, J. Mohtasham, R. W.
Winter, and G. L. Gard, Phys. Rev. B 65, 180506(R) (2002).

\bibitem{WIPRB2012} P. D. Grigoriev, M. V. Kartsovnik, W. Biberacher, Phys.
Rev. B \textbf{86}, 165125 (2012).

\bibitem{WIPRB2011} P.D. Grigoriev, Phys. Rev. B \textbf{83}, 245129 (2011).

\bibitem{WIPRB2013} P.D. Grigoriev, Phys. Rev. B \textbf{88}, 054415 (2013).

\bibitem{GG2014} A.D. Grigoriev and P.D. Grigoriev, Low Temp. Phys. \textbf{%
40}, 367 (2014) [Fiz. Nizk. Temp. 40(4), 472 (2014)]; arXiv:1310.7109v2.

\bibitem{SO} M.V. Kartsovnik, P.D. Grigoriev, W. Biberacher, N.D. Kushch, P.
Wyder, Phys. Rev. Lett. \textbf{89}, 126802 (2002).

\bibitem{Shub} P.D. Grigoriev, Phys. Rev. B \textbf{67}, 144401 (2003)
[arXiv:cond-mat/0204270].

\bibitem{PhSh} P.D. Grigoriev, M.V. Kartsovnik, W. Biberacher, N.D. Kushch,
P. Wyder, Phys. Rev. B \textbf{65}, 060403(R) (2002).

\bibitem{Abrik} A.A. Abrikosov, \textit{Fundamentals of the theory of metals}%
, North-Holland, 1988.

\bibitem{Shoenberg} Shoenberg D. \textquotedblright Magnetic oscillations in
metals\textquotedblright , Cambridge University Press 1984.

\bibitem{Ziman} J. M. Ziman, \textit{Principles of the Theory of Solids},
Cambridge Univ. Press 1972.

\bibitem{Mark88} M.V. Kartsovnik et al., Pis'ma Zh. Eksp. Teor. Fiz. \textbf{%
47}, 302 (1988) [JETP Lett. \textbf{47}, 363 (1988)].

\bibitem{SlowPRB2015} P.D. Grigoriev, A.A. Sinchenko, P. Lejay, A.
Hadj-Azzem, J. Balay, O. Leynaud, V.N. Zverev, and P. Monceau,
arXiv:1504.06064 [submitted to Phys. Rev. B].

\bibitem{LL10} E. M. Lifshitz and L. P. Pitaevskii, Course of Theoretical
Physics, Vol. 10: Physical Kinetics, (Nauka, Moscow, 2nd edition, 2002;
Pergamon Press, 1st edition, 1981).

\bibitem{LaTe08} N. Ru, R. A. Borzi, A. Rost, A. P. Mackenzie, J. Laverock,
S. B. Dugdale, and I. R. Fisher, Phys. Rev. B \textbf{78}, 045123 (2008).

\bibitem{Champel2001} V. M. Gvozdikov, Fiz. Tverd. Tela (Leningrad) 26, 2574
(1984) [Sov. Phys. Solid State 26, 1560 (1984)]; T. Champel and V. P.
Mineev, Phil. Magazine B \textbf{81}, 55 (2001).

\bibitem{ChampelMineev} T. Champel and V.P. Mineev, Phys. Rev. B \textbf{66}%
, 195111 (2002).

\bibitem{CommentDecay} For a short-range disorder the 2D electron wave
function in magnetic field decays exponentially at distance larger than the
Larmor radius.\cite{Fogler1997,Fogler1998} Therefore, for $\Delta y\gg R_{L}$
the matrix element $T_{mm^{\prime }}$ is exponentially small resulting from
the small overlap of the electron wave functions $\Psi _{m^{\prime }}^{\ast
}\left( r_{i}\right) \Psi _{m}\left( r_{i}\right) \sim \Psi _{m}^{\ast
}\left( r_{i}+\Delta y\right) \Psi _{m}\left( r_{i}\right) $.

\bibitem{Fogler1997} M. M. Fogler, A. Yu. Dobin, V. I. Perel, and B. I.
Shklovskii, Phys. Rev. B \textbf{56}, 6823 (1997).

\bibitem{Fogler1998} M. M. Fogler, A. Yu. Dobin, and B. I. Shklovskii, Phys.
Rev. B \textbf{57}, 4614 (1998).

\bibitem{Dingle} R.B. Dingle, Proc. Roy. Soc. \textbf{A211,} 517 (1952).

\bibitem{Bychkov} Yu. A. Bychkov, Zh. Exp. Theor. Phys. \textbf{39}, 1401
(1960), [Sov. Phys. JETP \textbf{12}, 977 (1961)].

\bibitem{AABook} B.L. Altshuler and A.G. Aronov "Electron-Electron
Interaction In Disordered Conductors", Ch. 1 in "Electron-Electron
Interactions in Disordered Systems", Ed. by A.L. Efros and M. Pollak,
Amsterdam: North-Holland (1985); ISBN: 978-0-444-86916-6.
\end{thebibliography}
\end{document}